\documentstyle[preprint,aps,epsf]{revtex}
\draft

\begin{document}
\baselineskip=24pt

\title{Monte Carlo Simulations of Conformal Theory Predictions
for the 3-state Potts and Ising Models}

\author{G.T. Barkema}

\address{Institute for Advanced Study, Princeton, NJ 08540}

\author{John McCabe\thanks{present address: mccabe@gwis2.circ.gwu.edu}}

\address{Laboratoire de Physique Th\'eorique ENSLAPP
\thanks{URA 14-36 du CNRS, associ\'e \`a l'Ecole Normale Sup\'erieure de
Lyon et \`a l'Universit\'e de Savoie.} \\
ENS Lyon, 46 all\'e d'Italie, 69364 Lyon Cedex 07, France}

\maketitle

\begin{abstract}
The critical properties of the 2D Ising and 3-state Potts models are
investigated using Monte Carlo simulations. Special interest is
given to measurement of 3-point correlation functions and
associated universal objects, i.e.  structure constants. The results
agree well with predictions coming from conformal field theory
confirming, for these examples, the correctness of the Coulomb
gas formalism and the bootstrap method.
\end{abstract}

\section{Introduction}

Conformal field theory \cite{1} has produced many precise predictions for
2-dimensional, equilibrium, critical systems. They fall into two large
groups: critical exponents \cite{2} and operator-product structure
constants \cite{3}. Theoretical calculations of these quantities are
based on very special properties of the representations of the conformal
group that are believed to be relevant to 2D critical statistical systems
\cite{4} -- degenerate Verma modules \cite{5} and modular invariance
\cite{6}. The calculation of structure constants exploits indirect
techniques of constructing correlation functions -- the bootstrap
equation \cite{5} and the Coulomb gas formalism \cite{3}. For many
important systems, like the Potts model, the only method available to
calculate these universal constants relies on conformal field theory.
Though there is widespread sentiment that these techniques are correct,
the abstractness of their analyses and the absence of independent
theoretical confirmation of their predictions justify an effort to
obtain experimental confirmation.

Experimental tests of the predictions for the structure constants
are very difficult and do not presently exist. They would
necessitate the measurement of both 2-point and 3-point correlation
functions at the critical point.  Such measurements are much simpler
in Monte Carlo simulations. The purpose of this article is to
report on Monte Carlo experiments for two well-known models
-- the Ising and 3-state Potts models. The results will give
``measured'' structure constants that will be compared to predictions
of conformal field theory \cite{8,9}. They will provide us with both an
``experimental'' test of conformal field theory and an insight into
the methods necessary to measure these new ``universal'' quantities
that it predicts in 2D critical systems.

We will briefly summarize some facts about the Potts and Ising models
that will be important to our analysis. A review of the statistical
properties of these models can be found in Refs. \cite{10,11}. Their
identifications with conformal field theories are discussed in Refs.
\cite{4,5,8,9}. Since the theoretical tools necessary to perform the
simulations are minimal, we refer to Refs. \cite{5,9} for explanations
of the theoretical calculation of structure constants.

Our simulations will be for models on square lattices with periodic
boundary conditions generated by two primitive vectors, $\vec{n}$.
The Hamiltonian of these models has the following form:
\begin{equation}
\label{ourhamil}
H= \sum_{\vec{x}} E(\vec{x}) =
 -J/2 \sum_{\vec{x},\vec{n}} \left[ S(\vec{x}+\vec{n}) +
S(\vec{x}-\vec{n}) \right] S^*(\vec x)
\end{equation}
where $\vec{x}$ is a vector on the lattice. The spin density, $S(\vec{x})$,
and the energy density, $E(\vec{x})$, are operators that
describe the coupling of the physical system to magnetic and
temperature perturbations respectively. They are fundamental
conformal operators (primary fields) having simple local definitions
on the lattice. The field $S(\vec{x})$ takes the discrete values $\pm 1$
for the Ising model and 1, $\exp(2\pi i/3)$, $\exp(-2\pi i/3)$ for the
3-state Potts model. The energy density operator is defined locally by
the value of $S(\vec x)$ on five neighboring lattice sites, i.e.
\begin{equation}
E(\vec{x})=-J \sum_{\vec{n}}
\frac{1}{2} \left[ S(\vec{x}+\vec{n})+S(\vec{x}-\vec{n}) \right]
S^*(\vec{x}). \end{equation}
Both of these operators exhibit scaling behavior at the critical point.

Traditionally, the Hamiltonian for the Potts model is written in a slightly
different form, i.e.
\begin{equation}
\label{tradhamil}
H= -J \sum_{\vec{x},\vec{n}} \delta (S(\vec{x}), S(\vec{x}+\vec{n})).
\end{equation}
The Hamiltonians (\ref{ourhamil}) and (\ref{tradhamil}) are equivalent,
except for an overall additive constant of $-J/2$ per bond, and a scaling
factor of 2/3: a pair of aligned neighboring spins contributes 1 to the
summation in both cases, but a pair of non-aligned neighboring spins
contributes -1/2 in (\ref{ourhamil}) and 0 in (\ref{tradhamil}).

The scaling behavior of 2-point correlations at a critical point
defines the conformal dimension, $\Delta_i$, of a scaling field
$\phi_i(\vec{x})$.  For spinless fields like $S(\vec{x})$ and
$E(\vec{x})$, it is given by:
\begin{equation}
\label{scaling}
\langle \phi_i (\vec{x}) {\phi^*_j}(\vec{0}) \rangle =
\frac{\delta_{ij} N_i^2} {|\vec{x}|^{4 \Delta_i}}
\end{equation}
True scaling fields (ex. conformal primary fields) have vanishing
statistical averages at a critical point,\cite{5} e.g.
$\langle \phi_i(\vec{x}) \rangle =0$. To obtain such
fields, one must subtract the thermal averages from lattice fields with
non-zero averages, like $E(\vec{x})$. Only the subtracted operators obey the
scaling law of (\ref{scaling}). The subtraction constants are not universal
and are not described by conformal theory. This subtraction procedure
must be explicitly done in any simulation.

Finally, we mention that (\ref{scaling}) and all other equations for critical
correlation functions manifestly respect the discrete symmetries of (1).
The spin field, $S(\vec{x})$, transforms under the discrete
symmetry $Z_2$ for the Ising and under $Z_3$ for the 3-state Potts model.
Its correlation functions will obey superselection rules, at the
critical point, associated with these symmetries.

The conformal dimensions of $S(\vec{x})$ and $E(\vec{x})$ have been known
for the Ising and Potts models for sometime \cite{4,5,9,10}. Their explicit
values are given in Table I.

Our main interest concerns the predictions from conformal field
theory for the 3-point correlation functions. It is well-known
that the 3-point correlations of conformally invariant theories
have the following special form \cite{1}:
\begin{equation}
\label{cft-scaling}
\langle \phi_i(\vec{x}_i) \phi_j(\vec{x}_j) \phi_k(\vec{x}_k) \rangle =
\frac{C_{ijk} N_i N_j N_k}
{|\vec{x}_{ij}|^{2\Delta_i + 2\Delta_j - 2\Delta_k} \times
{\mbox {\rm cyclic~perms.}}}
\end{equation}
The quantities $C_{ijk}$ are the structure constants. Much of the revival
of interest in conformal theories during the 1980's was associated
with the realization that, in 2D, the $C_{ijk}$'s were new universal
quantities different from critical exponents. More importantly,
they were shown to be calculable from symmetry considerations
alone \cite{5}. The $N_i$'s define the normalizations of the 2-point functions.
They are not universal and must be measured in our simulations of 2-point
correlations before extracting the universal constants,
$C_{ijk}$, from (\ref{cft-scaling}).

The calculation of the structure constants has been achieved for a
large variety of minimal conformal models by using their special
mathematical properties -- the existence of null vectors \cite{3,5}. These
models are believed to describe the critical behavior
of many of the important statistical systems. The critical point of the
Ising model has been identified with the $A_3$
conformal minimal model \cite{5}. The critical point of the 3-state
Potts model has been identified with a $Z_3$ symmetric version of the $D_5$
conformal minimal model \cite{4,9}. The values of the structure constants
resulting from theoretical calculations based on these identifications are
summarized in table \ref{structure}; the detailed calculations are found
in Refs. \cite{8,9}.

The structure constants not shown in table \ref{structure} vanish due to the
discrete symmetries of the two models. The vanishing of $C_{EEE}$ is,
however, less trivial. It results from a well-known fusion rule
($\Phi_{21}\times \Phi_{21} = 1$) whose discovery stimulated the revival of
interest in 2D conformal symmetry \cite{5} in 1984. We will compare this
prediction of conformal field theory directly to simulations. The fusion rules
only state when structure constants are non-zero. The actual values for
non-zero $C_{ijk}$'s of Table \ref{structure} have been calculated by using
two other tools of 2D conformal theory -- the
screened Coulomb gas formalism \cite{3} and the so called ``bootstrap''
equations \cite{5}. Thus, the values of the non-zero $C_{ijk}$'s are a
second fundamental prediction of conformal field theory. Experimental
confirmation of their explicit values supports the validity of these
latter two tools of 2D conformal theory.

While the constants of the Ising model were known before
the arrival of conformal field theories \cite{12}, those for the 3-state
Potts model have not been found by other methods \cite{9}. Thus, the later
model allows a real test of the methods of conformal field theory.

\section{Analysis}

Our simulations utilize the following procedure. First, the
infinite lattice critical temperature is found from duality
considerations $T_c=T_D$. Next, the exact value of $T_c$, for our finite
lattice, is determined by calculating the 2-point correlations of $S(\vec{x})$
near $T_D$: below $T_c$, the correlations approach a constant value at
large distances, above $T_c$, the correlations fall off exponentially,
and exactly at $T_c$ the correlations show power-law scaling behavior.
After constructing the 2-point correlation of $S(\vec{x})$ at $T_c$,
we can measure the scaling dimension of $S(\vec{x})$ and the
normalization constant $N_S$ in (\ref{scaling}). Next, the thermal
average of $E(\vec{x})$, $\langle E \rangle$, is measured at the critical
temperature. Then, the critical 2-point correlation of $E(\vec{x})$ is
simulated. The nonscaling contribution $\langle E \rangle^2$
is subtracted, and we then determine the scaling dimension of
$E(\vec{x})$ and $N_E$. Finally, the 3-point correlations are simulated.
The scaling exponents can be extracted and the structure constants are found
with the help of (\ref{cft-scaling}) and the values of $N_S$ and $N_E$.
The last step is to compare our ``simulated'' scaling dimensions and
structure constants with the conformal field theory predictions of
Tables \ref{dimensions} and \ref{structure}.

The Monte Carlo simulations of the Ising model were carried out on a
$512 \times 512$ square lattice with periodic boundary conditions.
The algorithm used to generate sample configurations is a
cluster algorithm, as outlined by Wolff \cite{wolff}. In one such cluster
move, the time scale was incremented by the fraction of spins included in
the cluster. In the work presented here, this time scale is only relevant
for defining the thermalization time and time between samples. In our Ising
simulations, we thermalized over 300 such time units and took 18000 samples
separated by 20 time units. To obtain statistical error bars, these samples
were blocked in groups of 150 samples, and their standard deviation was
obtained.

By determining the temperature at which the 2-spin correlation
$\langle S(\vec{x}) S(\vec{0}) \rangle$ shows power-law
scaling, we found that for our system size $T_c/T_D=1.0015 \pm 0.0002$.
All data reported in this work for the Ising model is obtained with
this lattice size and at the above value for $T_c$.
There are two independent theoretical calculations of $C_{SES}$ for
the Ising model, and one does not rely on conformal theory \cite{12}.
Here, we are sure of the value of $C_{SES}$ and the Ising model is a
test of our procedure. It gives us more
assurance when simulating the Potts model where no independent
checks exist. The simulations of the Potts model
are the real experimental tests of conformal field theory.

Fig. \ref{IsingSS} shows the 2-point correlations for
the Ising model. It is of particular importance to mention that the
vacuum expectation value, $\langle E \rangle^2$, has already been
subtracted in graphs of correlation functions of $E(\vec{x})$.
The measured exponents $\eta_{SS}=0.26 \pm 0.01$ and
$\eta_{EE}=2.01 \pm 0.05$ are in agreement with the values presented in
Table \ref{dimensions}, i.e. four times the conformal dimensions for each
field.
More importantly, Fig. \ref{IsingSS} tells us that
$N_S^2=0.704 \pm 0.004$ and $N_E^2=0.42 \pm 0.02$.

The 3-point correlations were measured by placing one operator at
$\vec{0}=(0,0)$ (the center of the lattice), and the two others at
$\vec{r}_1=(r,0)$ and $\vec{r}_2=(0,r)$, i.e. two vectors along the two
perpendicular lattice directions at a distance $r$. Fig. \ref{IsingSES}
shows this correlation for the Ising model as a function of $r$.
Conformal field theory, eq. (\ref{cft-scaling}) and Table \ref{dimensions},
predicts that
for the Ising model:
\begin{equation}
\langle S_{r_1} E_0 S_{r_2} \rangle =
2^{3/8} N_S^2 N_E C_{SES} r^{-5/4}.
\end{equation}
The observed power law behavior $\eta_{SES}=1.30 \pm 0.05$ is in
agreement with the predicted exponent. The prefactor $k_{SES}=0.33 \pm 0.02$
and the measured values of $N_S$ and $N_E$ allow us to calculate the structure
constant. We find that $C_{SES}=0.54 \pm 0.05$. This measured value
compares well with the theoretical value of Table \ref{structure}. Having shown
that our simulations work for the Ising model, we can proceed to
simulating the Potts model with some confidence.

For the 3-state Potts model, we repeat the analysis. The Monte Carlo
simulations were performed on a $500 \times 500$ lattice with the Wolff
algorithm. Again, we thermalized over 300 time units, and took samples
separated by 20 time units. For the Potts model, the total number of
samples was 21000. As for the Ising model, these samples were blocked in
groups of 150 to obtain statistical errors.

By determining the temperature at which the 2-spin correlation
$\langle S(\vec{x}) S^*(\vec{0}) \rangle$ shows power-law scaling,
we found that for our system size $T_c/T_D=1.0005 \pm 0.0003$.
The simulations of the 2-point functions give $N_S^2=0.54 \pm 0.03$
and $N_E^2=0.125 \pm 0.005$ (Fig. \ref{PottsSS}). The power-law dependence
of these correlations were measured as $\eta_{SS^*}=0.26 \pm 0.02$,
in agreement with Table \ref{dimensions}, and
($\eta_{SS^*}=1/4$), and  $\eta_{EE}=1.66 \pm 0.04$, in slight disagreement
with Table \ref{dimensions} ($\eta_{EE}=8/5$).
If we again place one operator at $\vec{0}=(0,0)$ (the center of the
lattice), and the two others at $\vec{r}_1=(r,0)$ and $\vec{r}_2=(0,r)$,
the values of Table \ref{dimensions} in combination with (\ref{cft-scaling})
predict that for the Potts model:
\begin{equation}
\langle S_{r_1} S_0 S_{r_2} \rangle =
2^{-1/15} N_S^3 C_{SSS} r^{-2/5}.
\end{equation}
and
\begin{equation}
\langle S_{r_1} E_0 S_{r_2}^* \rangle =
2^{4/15} N_S^2 N_E C_{SES^*} r^{-16/15}.
\end{equation}
The measurements shown in Fig. \ref{PottsSSS} fit this power-law behavior well:
$\langle S_{r_1} S_0 S_{r_2} \rangle=k_{SSS} r^{-\eta_{SSS}}$,
where $k_{SSS}=0.44 \pm 0.04$ and $\eta_{SSS}=0.39\pm 0.02$, and
$\langle S_{r_1} E_0 S^*_{r_2} \rangle=k_{SES^*} r^{-\eta_{SES^*}}$,
where $k_{SES^*}=0.14 \pm 0.01$ and $\eta_{SES^*}=1.11 \pm 0.04$,
respectively. The exponents agree with the theoretical values $2/5$ and
$16/15$. Combining these results with (\ref{cft-scaling}), we obtain measured
values for the two non-zero structure constants of the 3-state
Potts model, $C_{SES^*}=0.61 \pm 0.06$ and $C_{SSS}=1.16 \pm 0.14$.
Again, the agreement with the predictions of Table \ref{structure}
is quite good.

Finally, we show in Figs. \ref{IsingEEE}, \ref{PottsEEE} the simulations
of the 3-point correlation $\langle E_{r_1} E_0 E_{r_2} \rangle$.
For both the Ising and the Potts models, this
correlation should vanish due to the conformal fusion rule
$\Phi_{21}\times \Phi_{21} =1$ \cite{4,5}. As these graphs show, the
3-point correlation functions are consistent with zero when the distances
between the operators are as small as 5 lattice spacings (Note that the
scale of the graphs is greatly magnified.). At very
short distances the correlations do not vanish,
because the system sees the discreteness of the lattice and is thus
not described by conformal theory.
The vanishing of these correlations at distances greater than a few
lattice spacings is strong evidence for the correctness of the
conformal fusion rule.

In summary, we have measured 2- and 3-point correlations for the
two-dimensional Ising and 3-state Potts models and have compared both their
exponents and prefactors, e.g. structure constants, with predictions from
conformal field theory. All measured exponents, except one, are within one
standard deviation of theoretical predictions, and the remaining one is
within two standard deviations. Far more interesting is
the fact that the structure constants are also in good agreement with
conformal theory predictions. For $C_{EEE}$, this gives us a verification
of one fusion rule of conformal models. For $C_{SES^*}$ and $C_{SSS}$,
this gives a direct test of the screened Coulomb gas and bootstrap
equation formalisms which were necessary to obtain their theoretical
values. Our Monte Carlo simulations strongly support the
validity of the detailed conformal theory methods that have allowed the
calculation of higher correlation functions at critical points.

GTB acknowledges support from the DOE under grant DE-FG02-90ER40542 and the
Monell Foundation.
JM wishes to thank T. Wydro for helpful discussions and to also thank the
Laboratoire de Physique Th\'eorique of the ENS de Lyon for its hospitality
during initial stages of this work.

\newpage

\begin{table}[htb]
\begin{center}
\begin{minipage}{2.5in}
\caption{Scaling dimensions of the Spin and Energy Density Fields
\label {dimensions}}
\begin{tabular}{lcc}
\noalign{\smallskip}
Model: & Ising & 3-state Potts \\
\noalign{\smallskip}
\hline
$S(\vec{x})$ & 1/16 & 1/15 \\
$E(\vec{x})$ & 1/2 & 2/5 \\
\end{tabular}
\end{minipage}
\end{center}
\end{table}

\begin{table}[htb]
\begin{center}
\begin{minipage}{3.5in}
\caption{Predictions of structure constants from conformal theory
\label{structure}}
\begin{tabular}{lcc}
\noalign{\smallskip}
Model & Structure constant & value \\
\noalign{\smallskip}
\hline
Ising & $C_{SES}$ & 1/2 \\
 & $C_{EEE}$ & 0 \\
3-state Potts & $C_{SES^*}$ & 0.546 \\
 & $C_{EEE}$ & 0 \\
 & $C_{SSS} \left( =C_{S^*S^*S^*} \right) $ & 1.092 \\
\end{tabular}
\end{minipage}
\end{center}
\end{table}

\begin{figure}
\caption[]{2-spin correlations (top) and energy density 2-point correlations
(bottom) in the Ising model on a $512 \times 512$ lattice at $T=T_c$.
The dashed lines are our fit to the power-law behavior,
given by: $\langle S_0 S_r \rangle=k_{SS} r^{-\eta_{SS}}$,
where $k_{SS}=0.704 \pm 0.004$ and $\eta_{SS}=0.26 \pm 0.01$, and
$\langle E_0 E_r \rangle=k_{EE} r^{-\eta_{EE}}$,
where $k_{EE}=0.42 \pm 0.02$ and $\eta_{EE}=2.01 \pm 0.05$.
\label{IsingSS}}
\end{figure}

\begin{figure}
\caption[]{Mixed 3-point correlations of the Ising model on a $512 \times
512$ lattice at $T=T_c$. The dashed line is our fit to the
power-law behavior, given by:
$\langle S_{r_2} E_0 S_{r_2} \rangle=k_{SES} r^{-\eta_{SES}}$,
where $k_{SES}=0.33 \pm 0.02$ and $\eta_{SES}=1.30 \pm 0.05$.
\label{IsingSES}}
\end{figure}

\begin{figure}
\caption[]{2-spin correlations (top) and energy density 2-point correlations
(bottom) in the 3-states Potts model on a $500 \times 500$ lattice at $T=T_c$.
The dashed lines are our fit to the power-law behavior,
given by: $\langle S_0 S_r^* \rangle=k_{SS^*} r^{-\eta_{SS^*}}$, where
$k_{SS^*}=0.54 \pm 0.03$ and $\eta_{SS^*}=0.26 \pm 0.02$, and
$\langle E_0 E_r \rangle=k_{EE} r^{-\eta_{EE}}$,
where $k_{EE}=0.125 \pm 0.005$ and $\eta_{EE}=1.66 \pm 0.04$.
\label{PottsSS}}
\end{figure}

\begin{figure}
\caption[]{3-spin correlations (top) and mixed 3-point correlations (bottom)
in the Potts model on a $500 \times 500$ lattice at $T=T_c$.
The dashed lines are our fit to the power-law behavior,
given by: $\langle S_{r_1} S_0 S_{r_2} \rangle=k_{SSS} r^{-\eta_{SSS}}$,
where $k_{SSS}=0.44 \pm 0.04$ and $\eta_{SSS}=0.39\pm 0.02$, and
$\langle S_{r_1} E_0 S_{r_2}^* \rangle=k_{SES^*} r^{-\eta_{SES^*}}$,
where $k_{SES^*}=0.14 \pm 0.01$ and $\eta_{SES^*}=1.11\pm 0.04$.
\label{PottsSSS}}
\end{figure}

\begin{figure}
\caption[]{Energy-density 3-point correlations of the Ising model on a
$512 \times 512$ lattice at $T=T_c$. For $r>10$,
$\langle E_{r_1} E_0 E_{r_2} \rangle < 2\cdot 10^{-4}$.
\label{IsingEEE}}
\end{figure}

\begin{figure}
\caption[]{Energy-density 3-point correlations of the Potts model on a
$500 \times 500$ lattice at $T=T_c$. For $r>10$,
$\langle E_{r_1} E_0 E_{r_2} \rangle < 10^{-5}$.
\label{PottsEEE}}
\end{figure}

\end{document}